\documentclass[a4paper,11pt]{article}
\usepackage{jheppub} 
\usepackage{subfig}
\usepackage{siunitx}
\usepackage{graphicx}

\newcommand{\nubb}{0\nu\beta\beta}
\newcommand{\Kr}{^{83m}\text{Kr}}
\newcommand{\Xe}{^{136}\text{Xe}}

\title{\boldmath Measurement of Energy Resolution with the NEXT-White Silicon Photomultipliers}

\collaboration{NEXT collaboration}

\author[1,a]{T.~Contreras\note[a]{ Now at Fermi National Accelerator Laboratory},}
\author[2]{B.~Palmeiro,}
\author[2]{H.~Almaz\'an,}
\author[3]{A.~Para,}
\author[4]{G.~Mart\'inez-Lema,}
\author[2]{R.~Guenette,}
\author[5]{C.~Adams,}
\author[6]{V.~\'Alvarez,}
\author[7]{B.~Aparicio,}
\author[8]{A.I.~Aranburu,}
% ...
\affiliation[1]{
Department of Physics, Harvard University, Cambridge, MA 02138, USA}
\affiliation[2]{
Department of Physics and Astronomy, Manchester University, Manchester. M13 9PL, United Kingdom}
\affiliation[3]{
Fermi National Accelerator Laboratory, Batavia, IL 60510, USA}
\affiliation[4]{
Unit of Nuclear Engineering, Faculty of Engineering Sciences, Ben-Gurion University of the Negev, P.O.B. 653, Beer-Sheva, 8410501, Israel}
\affiliation[5]{
Argonne National Laboratory, Argonne, IL 60439, USA}
\affiliation[6]{
Instituto de Instrumentaci\'on para Imagen Molecular (I3M), Centro Mixto CSIC - Universitat Polit\`ecnica de Val\`encia, Camino de Vera s/n, Valencia, E-46022, Spain}
\affiliation[7]{
Department of Organic Chemistry I, University of the Basque Country (UPV/EHU), Centro de Innovaci\'on en Qu\'imica Avanzada (ORFEO-CINQA), San Sebasti\'an / Donostia, E-20018, Spain}
\affiliation[8]{
Donostia International Physics Center, BERC Basque Excellence Research Centre, Manuel de Lardizabal 4, San Sebasti\'an / Donostia, E-20018, Spain}
% ...
\emailAdd{tcontrer@fnal.gov}

\abstract{The NEXT-White detector, a high-pressure gaseous xenon time projection chamber, demonstrated the excellence of this technology for future neutrinoless double beta decay searches using photomultiplier tubes (PMTs) to measure energy and silicon photomultipliers (SiPMs) to extract topology information. This analysis uses $\Kr$ data from the NEXT-White detector to measure and understand the energy resolution that can be obtained with the SiPMs, rather than with PMTs. The energy resolution obtained of (10.9 $\pm$ 0.6) $\%$, full-width half-maximum, is slightly larger than predicted based on the photon statistics resulting from very low light detection coverage of the SiPM plane in the NEXT-White detector. The difference in the predicted and measured resolution is attributed to poor corrections, which are expected to be improved with larger statistics. Furthermore, the noise of the SiPMs is shown to not be a dominant factor in the energy resolution and may be negligible when noise subtraction is applied appropriately, for high-energy events or larger SiPM coverage detectors. These results, which are extrapolated to estimate the response of large coverage SiPM planes, are promising for the development of future, SiPM-only, readout planes that can offer imaging and achieve similar energy resolution to that previously demonstrated with PMTs.}

\begin{document}
\maketitle
\flushbottom


\section{Introduction}
\label{sec:background}

The nature of neutrinos, whether Majorana or Dirac, is still unknown. Neutrinoless double beta decay ($\nubb$) is the most promising method of determining if neutrinos are their own antiparticle, and many experiments have produced lower limits on the half-lives of this process for various isotopes \cite{KamLand-Zen2023,GERDA2020,NEXT:0vbb}. The goal of the next generation of experiments is to reach a sensitivity to the half-life of $\nubb$ decay ($T_{1/2}^{0\nu}$) covering the inverted neutrino mass ordering, which requires ultra low-background tonne-scale detectors with excellent energy resolution \cite{jones2022physics}. 

The Neutrino Experiment with a Xenon TPC (NEXT) program searches for $\nubb$ decay with a high-pressure gaseous xenon time projection chamber (TPC), using the double beta decay isotope $\Xe$. In the current generation of NEXT detectors, photomultiplier tubes (PMTs) are used to measure the primary scintillation light and total energy of events, while silicon photomultipliers (SiPMs), located behind an electroluminescence (EL) amplification region, are used to track the position and reconstruct the topology of events. NEXT-White demonstrated excellent energy resolution, with less than $1\%$ FWHM at \qty{2.5}{\mega\electronvolt} \cite{NEXT:Eres} (the region of interest) and $4.5\%$ FWHM at \qty{41.5}{\kilo\electronvolt} \cite{Next:krcal} (the energy of the $\Kr$ decays used for calibration). NEXT-White additionally showed a good understanding of backgrounds \cite{NEXT:bkgrds} while reducing them using event topology \cite{NEXT:topology,NEXT:DNN,NEXT:RL}. The detector also produced a measurement of the half-life of the double beta decay with neutrinos \cite{NEXT:2vbb} and a limit to the neutrinoless double beta decay \cite{NEXT:0vbb} for $\Xe$, consistent with other experiments.  

While the current NEXT-100 detector \cite{NEXT:N100},  a \qty{100}{\kilo\gram} scale detector, aims to demonstrate the scalability of the technology, efforts on the design of a future tonne-scale detector are ongoing \cite{NEXT:tonne}. For future detectors, the PMTs used to measure energy in the NEXT-White and NEXT-100 detectors present multiple limitations, including their sensitivity to high pressure and their large contribution to the background. A possible future detector design could use a tracking plane made out of SiPMs that could also measure energy \cite{NEXT:tonne}. This will likely require over tens of thousands of SiPM channels. While many experiments use or propose to use this scale of SiPM channels (T2K \cite{T2K2012}, LHCb \cite{LHCb2022}, CMS \cite{CMS2017}, DUNE \cite{DUNE2021}, JUNO-TAO \cite{JUNO2023}, DarkSide-20k \cite{DarkSide2018}), the requirements of these detectors vary due to their respective science goals. In the case of NEXT, the desired resolution is below $1\%$ FWHM in the region of interest at \SI{2.5}{\mega\electronvolt}, a level that has not yet been demonstrated with SiPMs. Additionally, NEXT is unique in the combination of low energy events, the use of EL amplification, and room temperature electronics and thus has unique requirements for the analysis of the SiPM waveforms. In this paper, NEXT-White data is analyzed to extract energy resolution with SiPMs and to investigate their possible limitations for energy measurement in future detector designs.


\section{The NEXT-White detector}

\begin{figure}[htbp!]
    \centering
    \includegraphics[width=.7\textwidth]{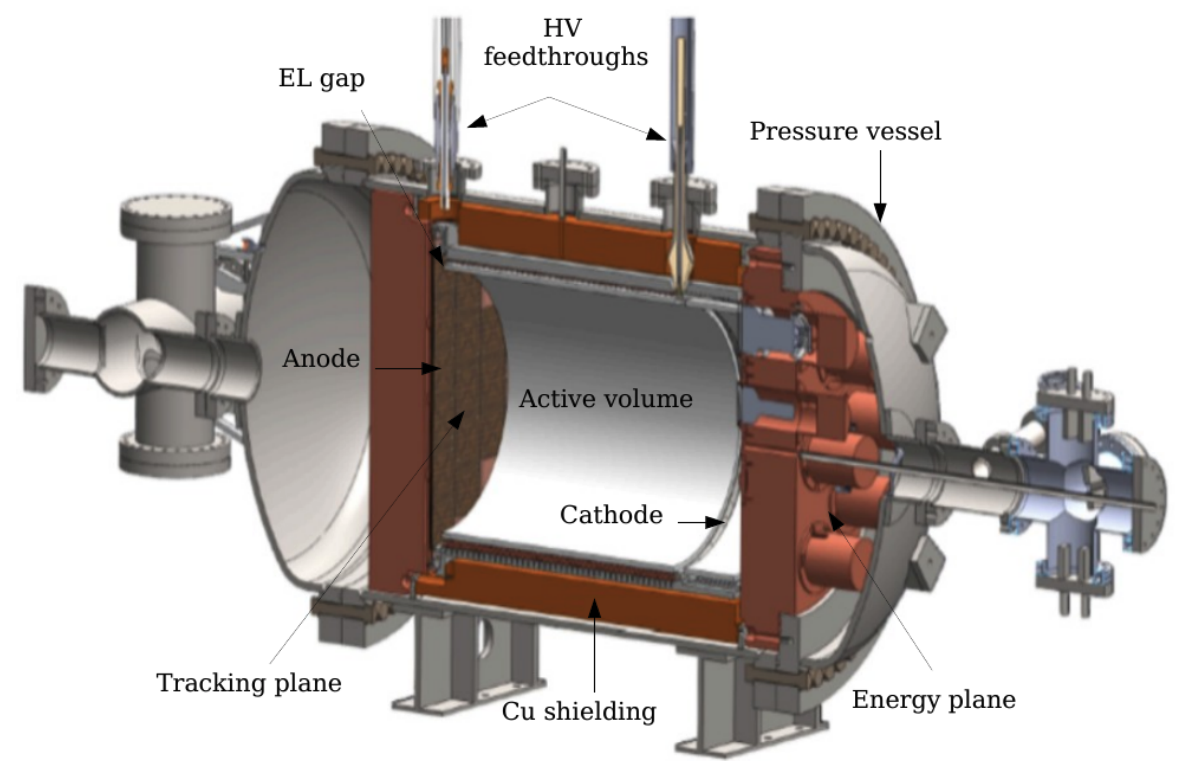}
    \caption{A diagram of the NEXT-White detector. The tracking plane, made of SiPMs, is shown and labeled on the left side while the energy plane, made of PMTs, is on the right side.}
    \label{fig:new}
\end{figure}

The NEXT-White detector, diagrammed in Figure~\ref{fig:new} and described in detail elsewhere \cite{NEXT:NEW}, is a high-pressure gaseous xenon time projection chamber. A particle interacting in the detector ionizes and excites the xenon gas, producing primary scintillation light, called S1, and primary ionization electrons. These electrons are drifted by the electric field toward the electroluminescence (EL) region, where the ionization electrons are accelerated to produce secondary scintillation light (S2). This light is collected by the SiPMs on the tracking plane, placed close behind the EL region, and by the PMTs on the energy plane, located on the opposite side of the detector. The granularity of the SiPMs tracking plane provides the $x-y$ information within the tracking plane, while the longitudinal distance ($z$) information comes from the time difference between the S1 and S2 signals and the known electron drift velocity. The center of the tracking plane is set as the origin in $x-y$, and $z=0$ is considered the anode of the EL region with positive values in the drift region.

The energy plane is made of 12 \textsc{Hamamatsu} R11410-10 3-inch PMTs, giving the plane a 31\% coverage for light collection. The tracking plane includes  1792 \textsc{SensL} series-C \qty{1}{\milli\meter^2} SiPMs placed at a pitch of \qty{10}{\milli\meter}, giving the plane a 1\% coverage. This pitch was optimized for tracking purposes and not for energy measurement, therefore the resulting light collection coverage is too low for a precise energy measurement, which is done by the PMTs on the energy plane. In this study, a detailed investigation is performed to measure the energy and understand and quantify the potential limitations of energy measurement with SiPMs. 

To calibrate the detector, a $^{83}$Rb source with an intensity of \qty{1}{\kilo\becquerel} is placed in the gas system. This source nuclei decays to $\Kr$, which is circulated in the gas and decays to the stable ground state of $^{83}$Kr at a rate of $\sim$~\qty{100}{\hertz}, however, due to limitations in the DAQ, the maximum event rate was \qty{10}{\hertz}. The $\Kr$ decays produce point-like events with a known energy of \qty{41.5}{\kilo\electronvolt} uniformly throughout the detector. These events can be used to map in detail the geometric variations as well as the loss of energy for events farther from the EL region, due to electron attachment to impurities in the gas \cite{Next:krcal}. This loss of electrons can be described by an exponential decay, with the time constant known as the \textit{electron lifetime}.

\section{Data acquisition}
The detection of $\Kr$ events is triggered when at least two of the three PMTs in the center ring of the energy plane \cite{NEXT:0vbb} record a S2 signal from 5000 to 50000 analog-to-digital counts (ADC), roughly corresponding to energies between \qty{15}{\kilo\electronvolt} and \qty{150}{\kilo\electronvolt}. The read-out window is \qty{1600}{\micro\second}, centered around the trigger time at \qty{800}{\micro\second}, for both PMTs and SiPMs. The sampling periods for the two sensors are \qty{25}{\nano\second} (PMTs) and \qty{1}{\micro\second} (SiPMs). The PMTs provide a precise measurement of the S1 time with their finer binning, and then are rebinned to \qty{1}{\micro\second} for the rest of the analysis. 

At the data acquisition level, the entire waveform for each PMT is recorded, while the waveforms for the SiPMs are typically reduced by an algorithm called {\it zero suppression} that minimizes the data size.  However, to allow the best possible understanding of an energy measurement from the SiPMs, this analysis uses a non-zero suppressed run, allowing for the use of the entire waveform for each SiPM. This run, labeled as run 8088, was taken starting July 11, 2020, for a total of 24.2 hours, with 694170 low-energy triggers. The conditions of this non-zero-suppressed run limit this analysis to the use of $\Kr$ events. 

\begin{figure}[htbp!]
\centering 
\includegraphics[width=.7\textwidth,clip]{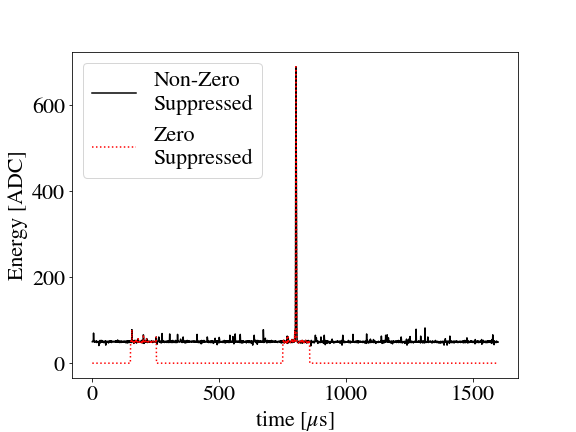}
\caption{\label{fig:zs} An example SiPM raw waveform (black) from run 8088 compared to an emulated zero suppressed waveform (red). The sharp peak is the S2 signal, while other smaller peaks are thermal and electronic fluctuations. In this example, the zero suppressed waveform successfully picks out the S2 region, in addition to a random region with high coincidence noise.}
\end{figure}


\section{Definition and contributions to the energy resolution}

In this paper, we define the relative energy resolution of a detector by the mean signal $\mu$ and its standard deviation $\sigma$ is given as:

\begin{equation}
\label{eq:eres}
    \frac{\delta E}{E} = 2~\sqrt{2 \ln{2}} ~\dfrac{\sigma}{\mu}\quad.
\end{equation}
For an ideal Gaussian response, this would correspond to the full width at half maximum (FWHM), the typical metric used to quantify detector energy resolutions in nuclear physics. For the NEXT-White detector, following \cite{NEXT:EresStudies} and assuming each contribution is independent of the others, the energy resolution can be written as:

\begin{equation} \label{eq:eres_all}
    \frac{\delta E}{E} =  2\sqrt{2\ln{2}} \sqrt{ \frac{F}{\overline{N}_i} + \frac{1}{\overline{N}_i}\frac{J}{\overline{N}_{EL}} + \frac{1}{\overline{N}_{PE}} + \frac{(\sigma_q/q)^2}{\overline{N}_{PE}} + \frac{\sigma_n^2(V)}{\overline{N}_{PE}^2}} \quad,
\end{equation} 

where $F$ is the Fano factor, $\overline{N}_i$ is the average number of ionization electrons produced, $J$ is the Conde-Policarpo factor or the relative variance in the number of photons produced in the EL region per ionization electron with an average of $\overline{N}_{EL}$ (i.e. the EL gain), $\overline{N}_{PE}$ is the average number of photons detected, $\sigma_q/q$ is the relative variance in the sensor gain, $\sigma_n^2(V)$ is the variance of the noise rate of the sensors, and $V$ represents the volume of the event being measured, or the 3D extent of the ionized electrons produced by the primary particle. This volume determines the number of SiPMs and times slices used in the measurement, which in the case of $\Kr$ events is an approximately constant volume. An increase in the effective detection volume must be considered for the extrapolation to higher energy events, as discussed in Section~\ref{sec:dis}.

The first term accounts for the variance in the number of ionization electrons, where, for gaseous xenon, $F = 0.15$ \cite{Fano1947}. The second term accounts for the variance in the amplification of the signal in the EL region. Measurements have shown that $J/\overline{N}_{EL} << F$, making this term negligible to the total energy resolution \cite{NEXT:EresStudies}. The third term represents the efficiency in collecting the photons, which can be approximated with a Poisson distribution for a large number of photons and low collection efficiency. This is an effective estimate, as $\Kr$ events produce on the order of $700,000$ photons and the number of photoelectrons measured is  $\sim 13,000$ with PMTs and $\sim 730$ with SiPMs. The fourth term gives the variance in converting the detected photons to photoelectrons. For SiPMs, $\sigma_q/q \approx 0.1$, extracted from sensor calibration runs, making this term also negligible. The last term incorporates the variance of the noise rate in an event due to the SiPMs. This analysis measures and subtracts the average amount of noise in the SiPMs, and measures the contribution $\sigma_n^2$ to the energy resolution. 

Given that the output of the SiPMs is expressed in photoelectrons (PE), and the absolute PE to keV scaling with respect to SiPMs has not been performed, the energy in this analysis is expressed in terms of photoelectrons collected ($N_{PE}$). The non-negligible contributions to the energy resolution can thus be written as:

\begin{equation} \label{eq:eres_pe}
    \frac{\delta E}{E} \simeq  2\sqrt{2\ln{2}}  \sqrt{ \frac{F}{\overline{N}_i} + \frac{1}{\overline{N}_{PE}} \left( 1 + \frac{\sigma_n^2(V)}{\overline{N}_{PE}} \right)}
\end{equation} 


\section{PMT and SiPM data processing}

In typical analyses from the NEXT Collaboration, the PMT and SiPM signals are used to extract very distinct information. The large-area PMTs are used to trigger the data readout and to measure the energy, while the sparse SiPM plane is used for tracking information. The $x$, $y$, and $z$ positions of events are necessary to correct the energy variations across the detector that are due to geometrical effects and electron lifetime. 

In this analysis, the SiPMs are further used to measure energy and to determine the limitations of a SiPM energy-tracking plane. Only events with exactly one S1 and one S2 are used in this analysis. To avoid any bias and to optimize the way data was treated, run 8088 was divided into two samples, randomly chosen in time. The first sample, making up 85\% of the run, was used to obtain the geometrical and lifetime corrections to the energy, called the \textit{calibration sample}, and the second sample (15\% of the run) was used for measurements, called the \textit{measurement sample}. 


\subsection{PMT data processing} \label{sec:pmt_proc}

\begin{figure}[htbp!]
\centering 
\subfloat[]{\includegraphics[width=.49\textwidth,origin=c]{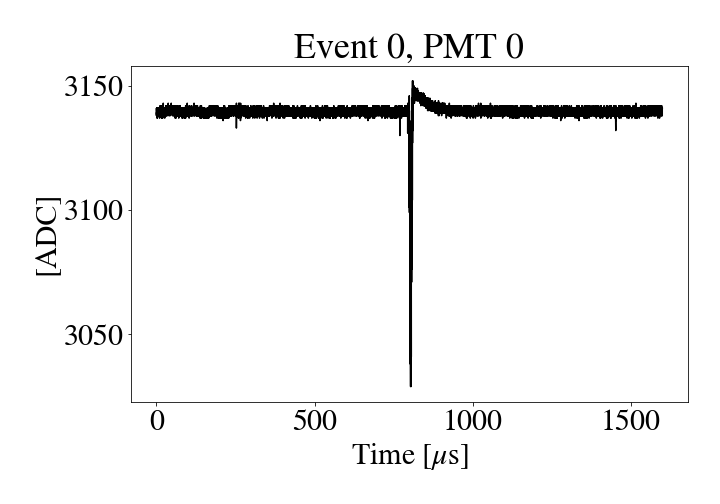}}
\subfloat[]{\includegraphics[width=.49\textwidth]{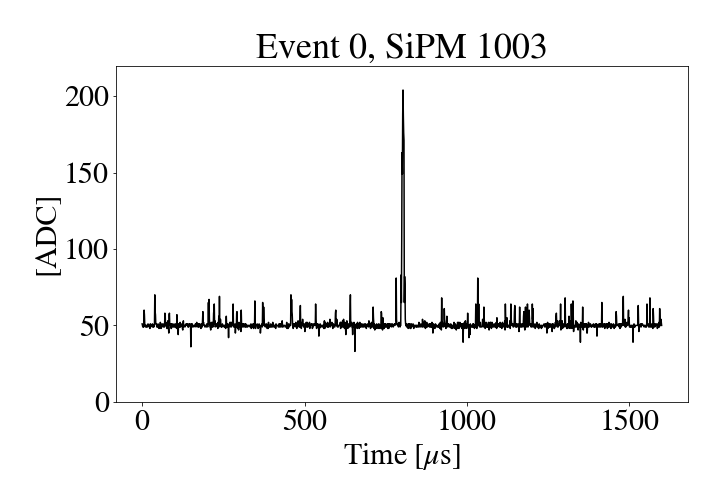}}
\centering
\caption{\label{fig:waveforms} (a) An example of a raw PMT waveform. The large peak represents the S2 signal. (b) An example raw non-zero suppressed SiPM waveform, including a non-zero baseline (around 50 ADCs) and electronic and thermal noise in addition to the signal.}
\end{figure}

The raw PMT waveforms, made of 64000 samples of \qty{25}{n\second} bins, have a non-zero baseline with a negative swing due to the AC coupling, as shown in Figure~\ref{fig:waveforms}(a). This is corrected using a deconvolution algorithm, producing positive-only, zero-baseline, calibrated waveforms \cite{NEXT:NEW}. A peak-finding algorithm is then used to identify S1 and S2 peaks. The width of the S2 peak varies depending on the drift length due to diffusion. The time difference between the S1 and S2 signal is used to determine the $z$ position of events, as the electron drift velocity is quite stable at 
\qty[per-mode = symbol]{1}{\milli\meter\per\micro\second} \cite{NEXT:drift}. To determine the energy of events using PMTs, allowing for direct comparison with the energy from SiPMs, the signal in the S2 time region is then summed for all PMTs. This energy is then corrected using a calibration \textit{map} made using the calibration sample, to correct for the geometric and electron lifetime effects, described in detail in Section~\ref{sec:corrections}.

\subsection{SiPM data processing} \label{sec:sipm_proc}

\subsubsection{SiPM noise subtraction}
\label{sec:sipm_noise}

\begin{figure}[htbp!]
\centering 
\includegraphics[width=.6\textwidth]{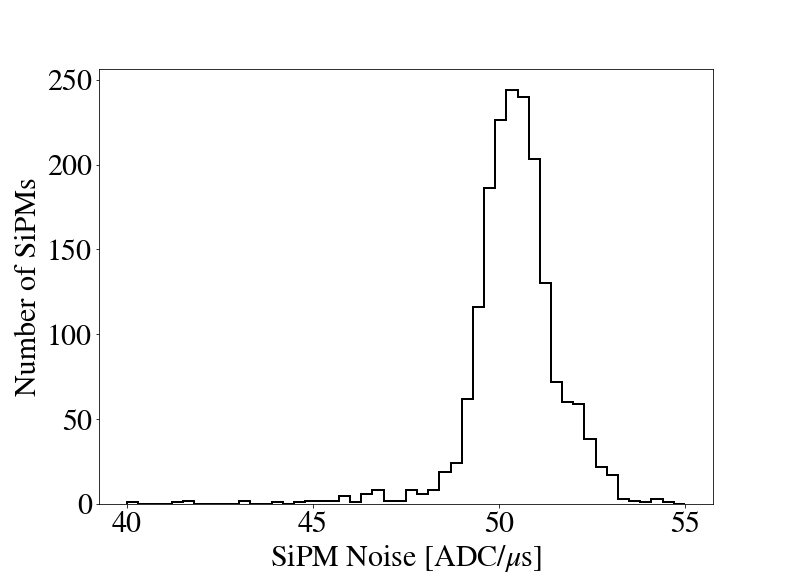}
\centering
\caption{\label{fig:baselines} The average signal per \qty{}{\micro\second} for each SiPM, excluding the \qty{20}{\micro\second}  near the S2 signal around \qty{800}{\micro\second}. This average is then subtracted from each sample to get the noise and baseline subtracted SiPM waveforms.}
\end{figure}

The SiPM waveforms are 1600 samples of \qty{1}{\micro\second} bins, as shown in Figure~\ref{fig:waveforms}(b) where the large peak around \qty{800}{\micro\second} shows the S2 signal. The triggered time, obtained from the PMTs, corresponds to the leading edge of a S2-like signal. The S1 for $\Kr$ events is too small to be seen with the NEXT-White SiPMs. This waveform includes a non-zero baseline set by the electronics, true photon signals, thermal pulses, crosstalk, after-pulsing, and the electronic noise introduced by the data acquisition system and front-end electronics. While the electronic fluctuations should be symmetric around the baseline and on average give zero when integrating over the electronic noise, the thermal pulses (dark counts), crosstalk, and after-pulsing will always be positive. As the SiPMs are at room temperature, the thermal pulses should dominate, however, in this analysis, we cannot distinguish between these noise sources. This results in a signal that increases with the length of the integration windows, just due to these noise sources. The mean of these fluctuations as well as the baseline must be removed so that when integrating to get an energy measurement, the signal shows an \textit{average of zero noise}. The average noise can be calculated for each SiPM using the average signal per \unit{\micro\second} bin for each SiPM, excluding the \qty{20}{\micro\second} region around the S2. The \qty{20}{\micro\second} exclusion region was chosen to remove any signal from the noise estimate, as the maximum width of the S2 signal has been measured as \qty{15}{\micro\second}. This average is then subtracted from the signal in each \qty{1}{\micro\second} bin, before any further processing. The resulting averages for each SiPM are shown in Figure~\ref{fig:baselines}, and an example waveform after the noise and baseline subtraction is shown in Figure~\ref{fig:windows}(a), where the S2 region is highlighted in blue and a region used to estimate the residual noise is highlighted in red. All SiPMs were shown to have a stable noise rate throughout run 8088.

After noise subtraction, the average residual noise in any \qty{1}{\micro\second} sample of a SiPM waveform will be zero, while the variance will be non-zero. This variance of the residual noise will contribute to the energy resolution of the detector and can be directly measured using the non-zero suppressed waveforms, as described in the following section.


\subsubsection{Event energy and tracking measurements with SiPMs}

\begin{figure}[htbp]
\centering 
\subfloat[]{\includegraphics[width=.465\textwidth]{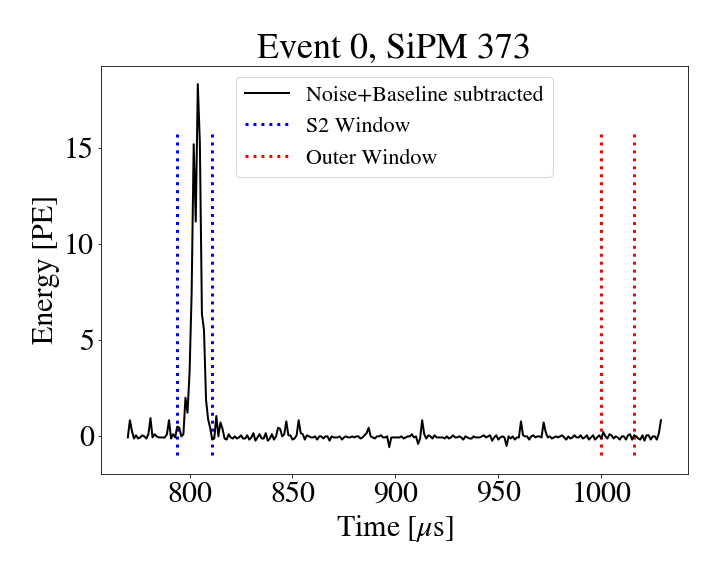}}
\subfloat[]{\includegraphics[width=.515\textwidth]{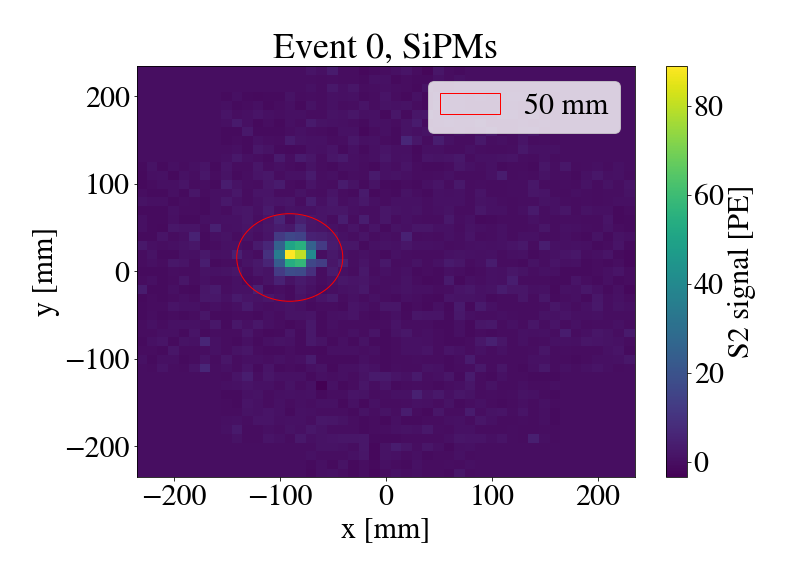}}
\caption{\label{fig:windows} (a) An example of partial waveforms after noise and baseline subtraction of one SiPM used to measure energy in this event. Highlighted in blue is the S2 window to measure event energy, and in red is the outer window used to estimate the variance of the noise, which, regardless of the noise subtraction, contributes to the energy resolution. (b) The summed signal over the S2 window of each SiPM for an event. The red circle encloses the SiPMs used for the energy measurement, and is centered around the reconstructed barycenter of the event.}
\end{figure}

\begin{figure}[htbp!]
\centering 
\includegraphics[width=.7\textwidth,clip]{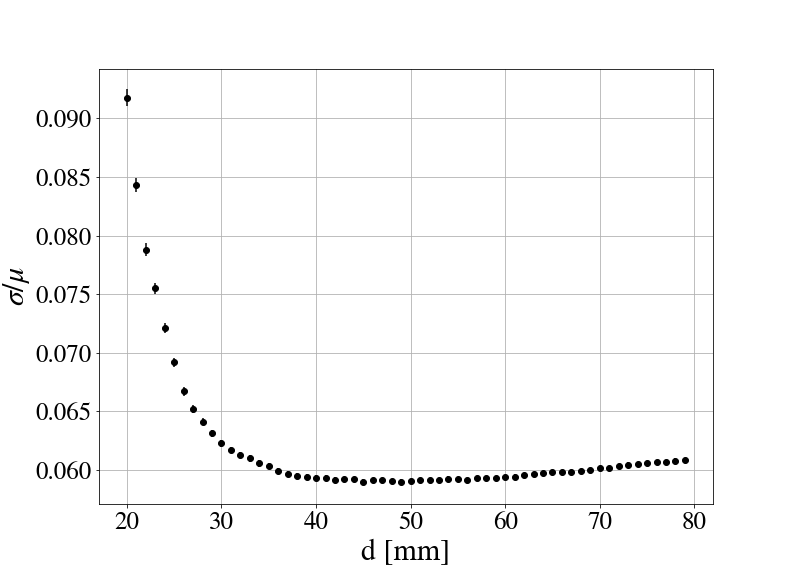}
\caption{\label{fig:dcuts} The standard deviation over the mean of the SiPM event energy using all SiPMs inside the distance $d$ from the event center. Error bars correspond to the statistical fluctuation.}
\end{figure}

After subtracting the average noise per SiPM, the signal in the S2 region is time-integrated for each SiPM. The event center is extracted based on a barycenter algorithm. A new algorithm is used here to measure the energy with SiPMs, using only SiPMs within a certain distance ($d$) of the event center. For all these SiPMs within $d$, the entire waveform inside the S2 region is kept, allowing SiPMs with smaller signals to contribute to the total. The optimal distance ${d = \qty{50}{\milli\meter}}$ from the event center was found by minimizing the $\sigma/\mu$ of the energy distribution, as shown in Figure~\ref{fig:dcuts}. This method is referred to as the \textit{SiPM distance method}. An example of the summed signal for all SiPMs is shown in Figure~\ref{fig:windows}(b), where only the signal within the red circle of ${d = \qty{50}{\milli\meter}}$ is used for the energy measurement.

In addition to calculating the energy in the S2 region, the residual noise after noise subtraction in each event is estimated. The distribution of this residual noise for many events has a mean of zero, as demonstrated later in this analysis (Figure~\ref{fig:noise}). However, the distribution has a non-zero variance, ultimately contributing to the variance of the energy distribution across events. This is done by selecting a time window far from the S2 window, as shown in Figure~\ref{fig:windows}(a), with the same length as the S2 window. This region is expected to include only noise and no true signal. The same SiPMs chosen to measure energy (using the SiPM distance method with $d=\qty{50}{\milli\meter}$) are used to estimate the noise. The waveforms within this outer window, for all chosen SiPMs, are summed to estimate the residual noise in the energy measurement. 


\subsection{Correction Maps}
\label{sec:corrections}

Two detector effects distort the energy of events and must be corrected. The first is the energy dependence on $z$, due to the electron attachment to impurities in the gas, characterized by an exponential decay with a decay constant known as the electron lifetime. The second is the $x-y$ dependence of energy due to the inhomogeneous light collection efficiency, particularly near the edges of the detector and from the suspected small tilt in the EL planes \cite{haefner_thesis}. Krypton calibration runs can be used to create detailed calibration \textit{maps} that help correct for both effects \cite{Next:krcal}.
	
\begin{figure}[htbp]
\centering 
\subfloat[]{\includegraphics[width=.45\textwidth]{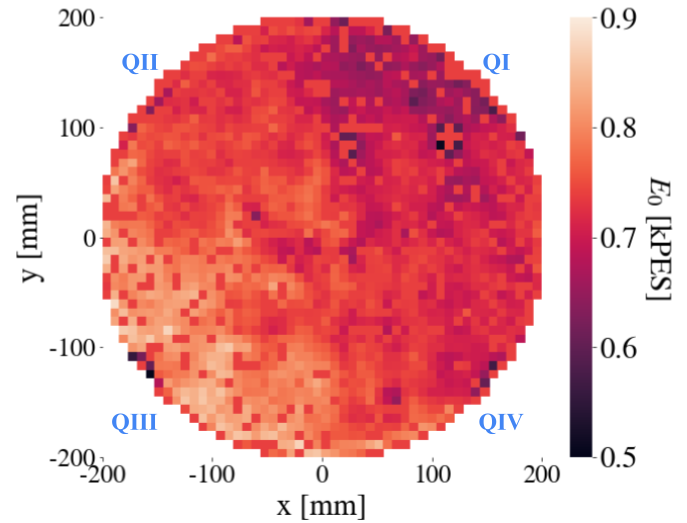}}
\subfloat[]{\includegraphics[width=.53\textwidth]{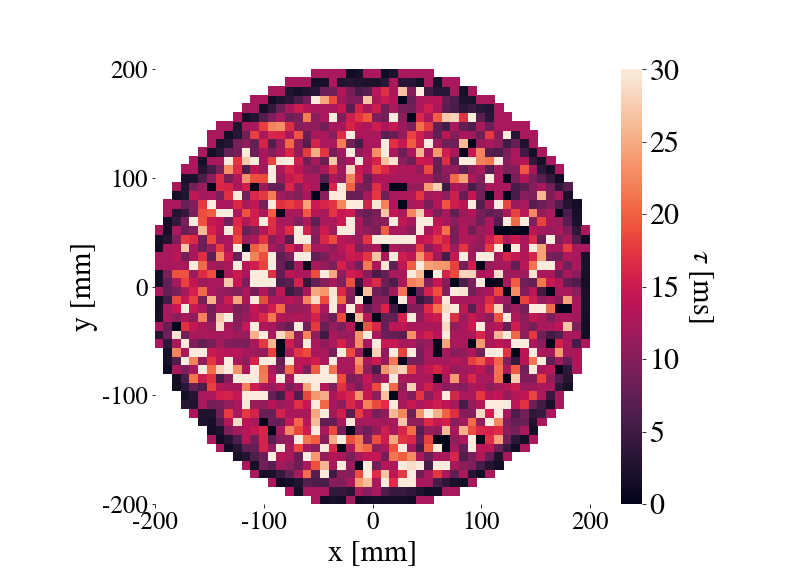}}
\caption{\label{fig:krmaps} Corrections maps using the calibration sample of run 8088 extracted from SiPMs signal, based on an exponential decay with (a) an initial energy at $z=0$ of $E_0$ and (b) lifetime $\tau$. Each quadrant of the detector is labeled in blue in (a).}
\end{figure}

The calibration sample of run 8088 is used to produce the calibration maps for both PMTs and SiPMs. Events are binned in $x-y$. The event energy versus the event $z$ position is fit for each $x-y$ bin to an exponential, ${E(z) = E_0\cdot \exp{(-z/\tau)}}$, where $E_0$ is the energy of events at $z=0$, and $\tau$ is the electron lifetime. A map based on the SiPM signal of the $E_0$ parameters for each $x-y$ bin in Figure~\ref{fig:krmaps}(a) represents the geometric effects, while the map of the $\tau$ in each $x-y$ bin in Figure~\ref{fig:krmaps}(b) represents the electron lifetime. The corresponding map using the PMT signal is shown in Appendix~\ref{sec:pmts} for reference.

\begin{figure}[htbp!]
\centering 
\subfloat[PMTs]{\includegraphics[width=.49\textwidth]{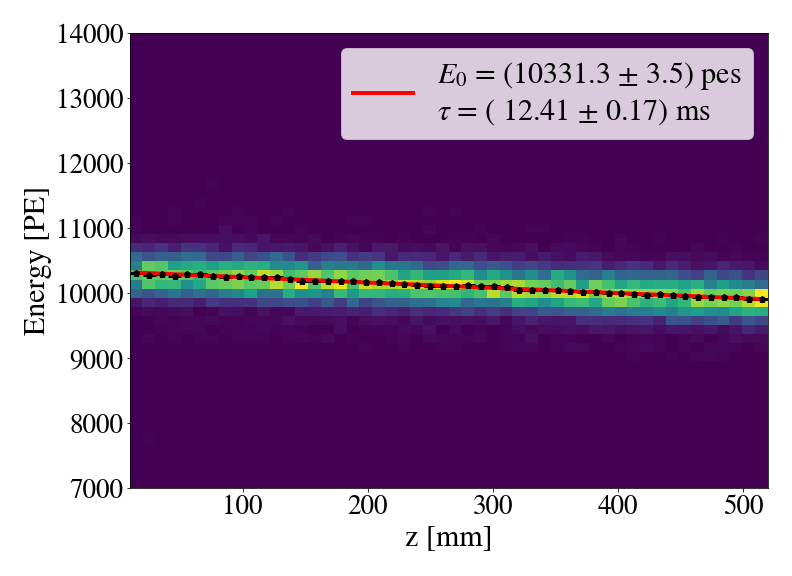}}
\subfloat[SiPMs]{\includegraphics[width=.49\textwidth,origin=c]{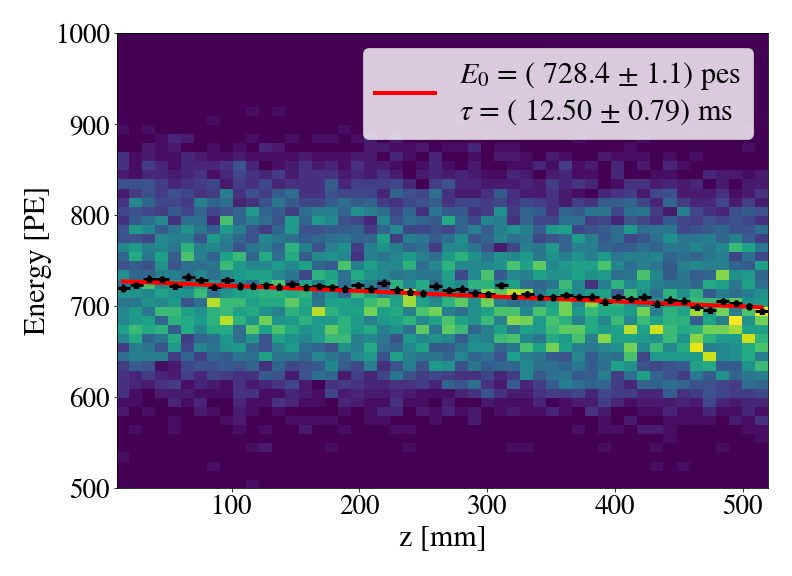}}
\centering
\caption{\label{fig:lifetime} The event energy as a function of the $z$ position, showing the decrease in energy with higher $z$ due to electron attachment, described by the electron lifetime $\tau$. This lifetime was measured using the signal from (a) the PMTs and (b) the SiPMs, with the SiPM distance method applied with a cut of ${d=\qty{50}{\milli\meter}}$. Both the PMT and SiPM signal use a fiducial radial cut of ${r<\qty{198}{\milli\meter}}$. The fit values for both the initial event energy ($E_0$) and for the lifetime ($\tau$) are also shown.}
\end{figure}

It is expected that the geometric effects will look different for the PMTs compared to the SiPMs. The PMTs are farthest from the EL gap, and thus see large effects from events at the edges of the detector, where more light would be lost (see for example Appendix~\ref{sec:pmts} Figure~\ref{fig:r_eres_pmt}). The SiPMs have a much flatter geometric distribution, as they are very close to the EL region, having consequently less drastic edge effects. Dead SiPMs can be seen as the three small dark regions in Figure~\ref{fig:krmaps}(a). However, the energy dependence on $z$, shown in Figure~\ref{fig:lifetime}, is expected to be consistent between the PMTs and SiPMs. This analysis, using the processing described above, results in an electron lifetime consistent with the one obtained with the traditional PMT analysis. 

\section{Results for energy resolution with SiPMs}

For this analysis, a fiducial volume cut excluding events with ${r<\qty{198}{\milli\meter}}$ and ${\qty{10}{\milli\meter}<z}$ ${<\qty{520}{\milli\meter}}$, and a cut excluding events within a distance of $\qty{15}{\milli\meter}$ from a dead or high variance SiPM (identified during sensor calibrations) are used. All SiPM energy measurements were made using the SiPM distance method with $d=\qty{50}{\milli\meter}$.

We determine the FWHM of the relative energy resolution from the mean and standard deviation of the measured distribution fit to a Gaussian, as in Eq.~\ref{eq:eres}. The energy resolution can vary across the detector as a result of geometric effects and electron lifetime in the detector. This causes the energy resolution to have both a radial and longitudinal dependency. Additionally, there is a difference in energy resolution across different quadrants in the detector, as expected from the tilt in the EL planes. This analysis compares these energy resolution dependencies after corrections. For comparison, the energy resolution using the PMT signal is also shown in Appendix~\ref{sec:pmts}.


\subsection{Study of the radial dependence} \label{sec:eres_r}

\begin{figure}[htbp]
\centering 
\includegraphics[width=.7\textwidth,clip]{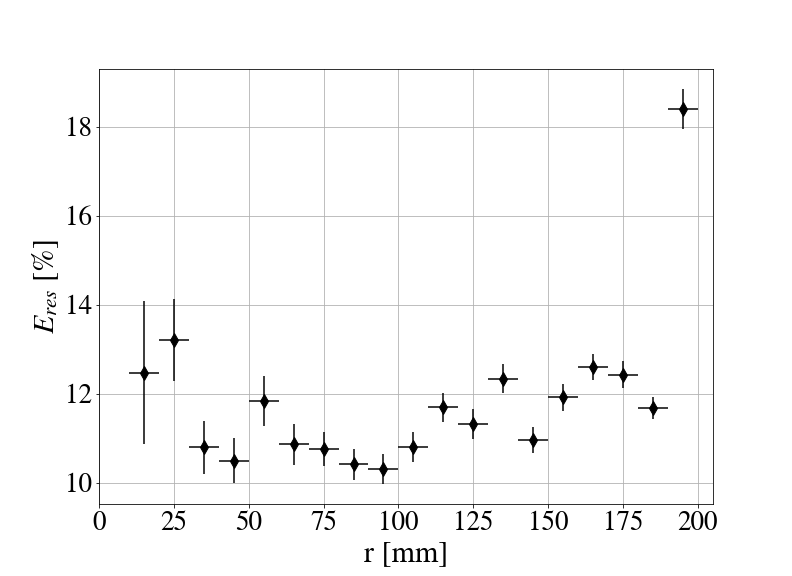}
\caption{\label{fig:r_vs_eres} Energy resolution (FWHM) using SiPMs as a function of the radial distance after corrections. Only events with ${\qty{50}{\milli\meter}<z<\qty{200}{\milli\meter}}$ are included, to remove any longitudinal dependencies. The errors correspond to statistical fluctuations.}
\end{figure}

Figure~\ref{fig:r_vs_eres} shows the energy resolution as a function of $r$ after corrections. To remove any longitudinal dependencies, only events with ${\qty{50}{\milli\meter}<z<\qty{200}{\milli\meter}}$ are included. The corrected results indicate that the SiPMs do not see a strong dependence of the energy resolution as a function of $r$, in contrast to PMTs that have a clear radial dependence (as seen in Figure~\ref{fig:r_eres_pmt}). This is explained by the fact that the SiPM plane is close to the EL gap, significantly reducing the geometrical variations away from the center of the detector.  A slightly worse energy resolution at very high radii, starting at approximately \qty{185}{\milli\meter} is observed and is attributed to events close to the edge of the detector.

\subsection{Study of longitudinal dependence} \label{sec:eres_z}

\begin{figure}[htbp]
\centering 
\includegraphics[width=.7\textwidth,clip]{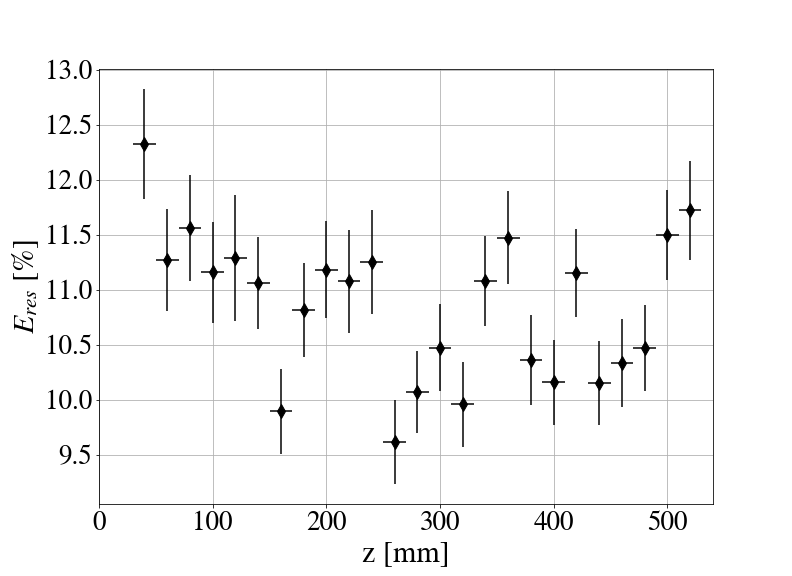}
\caption{\label{fig:z_vs_eres} Energy resolution (FWHM) using SiPMs as a function of the z position of the events after corrections. Only events with ${r<\qty{100}{\milli\meter}}$ are included, to remove radial dependencies. The error bars correspond to the statistical fluctuations.}
\end{figure}

The energy resolution as a function of the longitudinal distance $z$ using SiPMs after correction is shown in Figure~\ref{fig:z_vs_eres}. To remove any radial dependence, only events with ${r<\qty{100}{\milli\meter}}$ are included. For $z \gtrsim \qty{50}{\milli\meter}$, the corrected energy resolution is fairly stable. At low drift distances ($z \lesssim \qty{50}{\milli\meter}$) a slightly worse energy resolution is obtained. Detailed studies have shown that this dependence can be attributed to the large pitch and low coverage of the SiPMs in the NEXT-White detector. At very low z, the krypton events have the smallest electron cloud possible, since they have not diffused in the gas while drifting. The exact position of these krypton events in relation to the nearest SiPM greatly changes the amount of energy collected.  This geometrical effect was verified with Monte Carlo simulations. For krypton events at larger z, the electron cloud has diffused enough to spread across more SiPMs, lowering the variance in energy measurement. This is an interesting result of the impact of the low coverage of the NEXT-White SiPMs, and would be greatly improved with higher coverage. 

\section{Energy resolution in quadrants} \label{sec:quadrants}

\begin{figure}
    \centering
    \includegraphics[clip,width=1\linewidth]{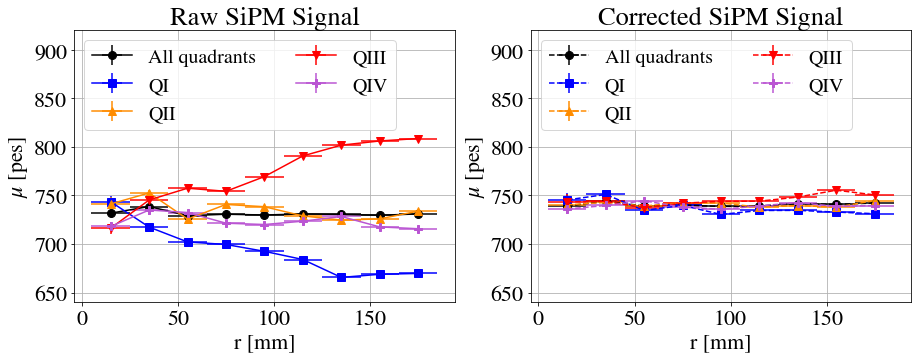}
    \caption{The radial dependence of the mean signal before corrections on the left, and after corrections on the right. The black line shows the mean using the entire detector, while the blue, yellow, red, and purple lines use only respectively quadrant I, II, III, and IV. The difference in gain between the top right versus the bottom left of the detector can be seen by the trends at large $r$ before corrections in QI and QIII. Events included are between $\qty{50}{\milli\meter}<z<\qty{200}{\milli\meter}$}
    \label{fig:r_mu_quad}
    \centering
\end{figure}

The tilt in the EL gap of NEXT-White, as measured indirectly in \cite{haefner_thesis}, has a much more prominent effect on the energy response of the SiPMs compared to the PMTs. This is due to the small distance of \qty{8}{\milli\meter} between the center of the EL region and the tracking plane. The effect of the EL gap can be seen by looking at the response in different quadrants of the detector, as labeled in Figure~\ref{fig:krmaps}(a), with the positive x and positive y sector (top right) labeled as quadrant 1 (QI), and QII, QIII, and QIV following sequentially counterclockwise around the origin. The uncorrected (left) and corrected (right) response in Figure~\ref{fig:r_mu_quad} shows the difference in gain between QI (upper right of Figure~\ref{fig:krmaps}(a)) and QIII (lower left of Figure~\ref{fig:krmaps}(a)). This difference in gain causes a large difference in photons per electron produced, leading to a larger variance in energy and ultimately a worse energy resolution when combining areas with different gain. There are also some remaining non-uniformities after correcting the energy, which could be improved with larger run statistics. 

\begin{figure}
    \centering
    \includegraphics[width=.7\linewidth]{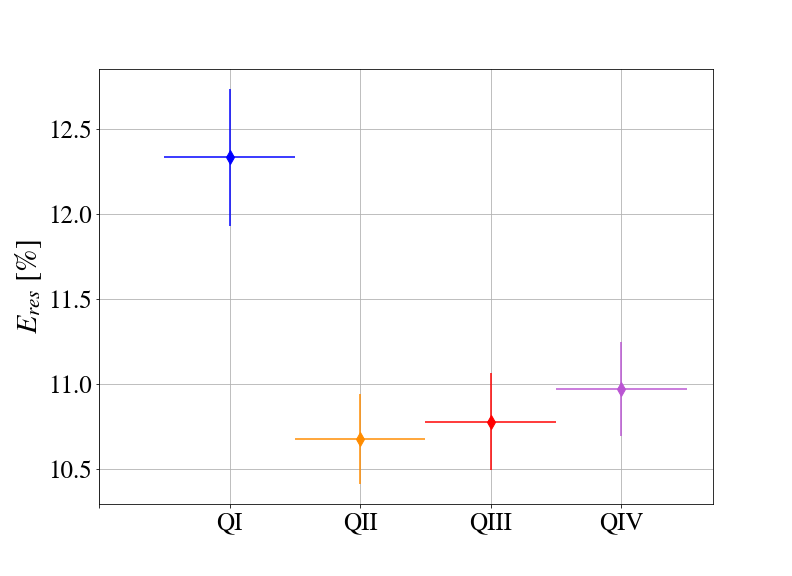}
    \caption{The energy resolution (FWHM) for each quadrant of the detector using the signal from the SiPMs after corrections. Events used have $r<198$ mm and $\qty{50}{\milli\meter}<z<\qty{520}{\milli\meter}$. The colors are used to match Figure~\ref{fig:r_mu_quad}.}
    \label{fig:eres_quads}
\end{figure}

Figure~\ref{fig:eres_quads} shows the energy resolution of each quadrant after corrections. QI is significantly worse than the others. This is likely due to the remaining non-uniformities seen in the corrected mean energy of Figure~\ref{fig:r_mu_quad}(b) as well as QI including the most dead SiPMs.

\subsection{Final Energy Resolution}

\begin{figure}[htbp]
\centering 
\subfloat[]{\includegraphics[width=.49\textwidth]{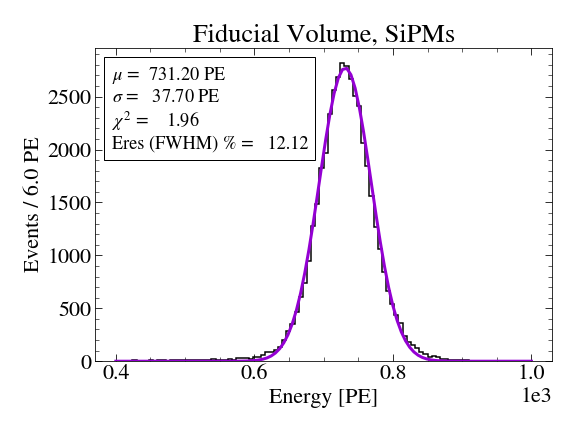}}
\subfloat[]{\includegraphics[width=.49\textwidth,origin=c]{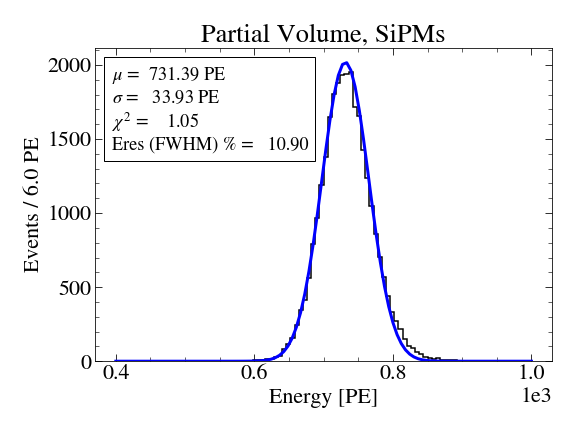}}
\centering
\caption{\label{fig:eres} The energy distribution using SiPMs using the (a) full detector, with a fiducial cut of $r<\qty{198}{\milli\meter}$ and $\qty{10}{\milli\meter}<z<\qty{520}{\milli\meter}$ and (b) partial detector, with cuts $r < \qty{185}{\milli\meter}$, $\qty{50}{\milli\meter} < z < \qty{520}{\milli\meter}$, and excluding quadrant QI. The energy resolution is calculated using Eq.~\ref{eq:eres} and a Gaussian fit of the distribution is used to extract $\mu$ and $\sigma$. The $\chi^2$ shown is the reduced chi-squared.}
\end{figure}

The systematic uncertainties were found by measuring the variance of the energy resolution when Gaussianly varying the correction maps, fit range, and binning, and were found to be consistent with previous studies \cite{Next:krcal}. Using the full detector volume after fiducial cuts, and applying the energy corrections described in sections~\ref{sec:corrections}, an energy resolution using SiPMs of (12.1~$\pm$~0.6) \% (FWHM) at \qty{41.5}{\kilo\electronvolt} is obtained. Since there are non-uniformities of the amount of light seen across the SiPM plane due to the EL tilt and given that the energy resolution is sensitive to the amount of light collected, the resolution fluctuates across the detector. To estimate the best possible energy resolution, we repeated the analysis using only events in the most homogenous regions of the detector. This includes ${r < \qty{185}{\milli\meter}}$ due to large variations in the event energy at the edges of the detector, ${\qty{50}{\milli\meter} < z < \qty{520}{\milli\meter}}$ due to detector edges and the variations at very low $z$, and QII-IV due to the EL plane tilt and dead SiPMs causing QI to have large variance in energy. These exclusions result in the partial volume defined by ${r < \qty{185}{\milli\meter}}$, ${\qty{50}{\milli\meter} < z < \qty{520}{\milli\meter}}$, and excluding quadrant QI. With this partial volume (resulting in $\sim52\%$ of the full detector volume), an energy resolution of (10.9 $\pm$ 0.6) $\%$ (FWHM) is achieved. The energy distributions and Gaussian fits for both the full and the partial volumes are shown in Figure~\ref{fig:eres}.

\begin{figure}[htbp]
\centering 
\includegraphics[width=.6\textwidth,clip]{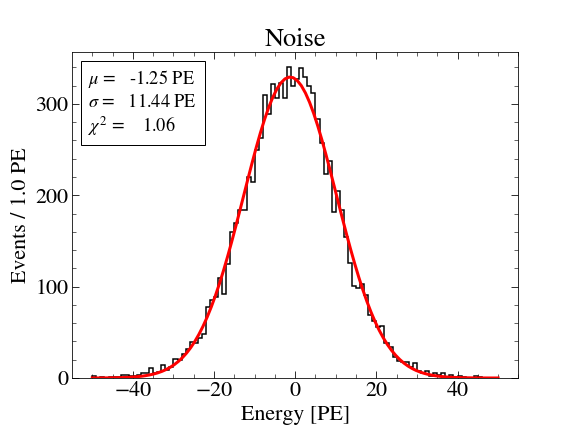}
\caption{\label{fig:noise} The distribution of measured photoelectrons of the same events used in the partial volume distribution in Figure~\ref{fig:eres}(b), using the samples outside S2. By fitting this distribution to a Gaussian, the fitted variance represents the expected noise contribution to the energy resolution of the partial volume distribution. $\chi^2$ is the reduced chi-squared.}
\end{figure}

The noise contribution to the energy resolution can be measured by estimating the residual noise in each event after noise subtraction, as described in Section~\ref{sec:sipm_noise}. Using the same events as used in the partial volume after cuts, the distribution of this residual noise is shown in Figure~\ref{fig:noise}. This noise distribution has a mean of (-1.3 $\pm$ 0.1) PE, close to zero as expected after noise subtraction, and a standard deviation of (11.44 $\pm$ 0.09) PE. This standard deviation directly contributes to the total shown in the energy distribution of the partial volume in Figure~\ref{fig:eres} (b) of ($33.9 \pm 0.3$) PE. 

\subsection{Estimation and extrapolation of energy resolution for higher coverage} \label{sec:extrap}

The expected energy resolution, using the signal achieved by the SiPMs, can be calculated based on Eq.~\ref{eq:eres_pe}. In the case of the partial detector volume, a mean energy of $N_{PE}~=~(731.4~\pm~0.3)$~PE, and a standard deviation of the noise of $\sigma_n~=~(11.44~\pm~0.09)$~PE are obtained. By applying these numbers to Eq.~\ref{eq:eres_pe}, an energy resolution of (9.68~$\pm$~0.01)~$\%$ (FWHM) is calculated. There is a small difference between this estimate and the best-achieved resolution in data (10.9 $\pm$ 0.6) $\%$. This discrepancy is most likely due to the limited non-zero suppressed data available, and the difficulties correcting geometric effects with the low coverage of the NEXT-White tracking plane. 

\begin{figure} 
    \centering
    \includegraphics[width=.6\linewidth]{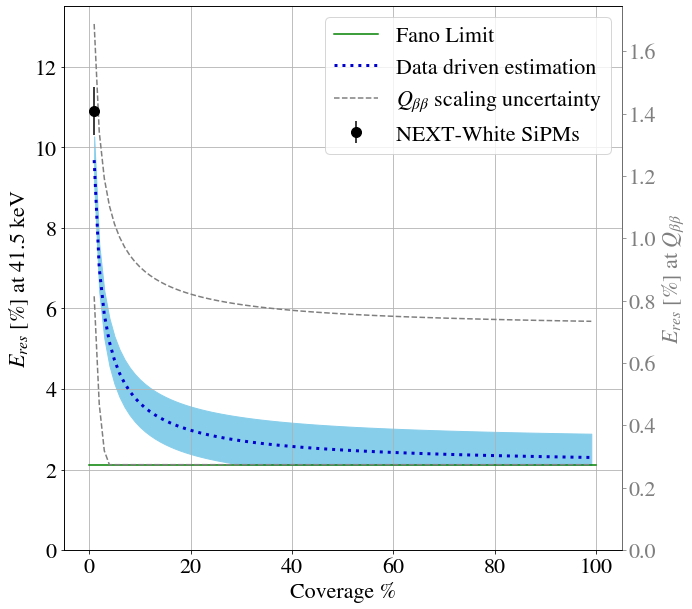}
    \caption{The data-driven estimated energy resolution versus coverage, assuming a NEXT-White geometry and SiPMs. The data-driven energy resolution is derived from $\Kr$ decays (left axis), and a naive $1/\sqrt{E}$ scaling to the $Q_{\beta\beta}$ value for $\Xe$ is shown on the right axis. The shaded blue region is the error of the estimate, accounting for the systematic error, the extrapolated statistical error of measured energy resolution. The grey dashed lines additionally include the error for scaling the energy resolution to $Q_{\beta\beta}$. The Fano limit is given by the green horizontal dotted line, and the energy resolution for the NEXT-White SiPMs using the partial volume is in black.}
    \label{fig:est_eres}
\end{figure}

The main contribution to the energy resolution is the photon statistics term. This is attributed to the low photon collection efficiency of the NEXT-White SiPMs due to their low coverage ($\sim 1 \%$). The expected energy resolution as a function of SiPM coverage for detectors like NEXT-White can be estimated by considering how each factor in Eq.~\ref{eq:eres_pe} will change as a function of the coverage. The total number of photons collected by the SiPMs will be linearly proportional to the coverage, while the Fano contribution will not change. The variance in the noise due to SiPMs will also increase linearly with coverage as the number of SiPMs increases. Using these relations and the results from NEXT-White SiPMs, the estimated energy resolution as a function of coverage is calculated and shown in Figure~\ref{fig:est_eres}. The colored blue region represents the estimated error based on the energy resolution with krypton, which includes the systematic error found with the partial volume, and the statistical error based on the number of photons collected (propagated to the given coverage). An additional uncertainty is included to account for the uncertainty in the $1/\sqrt{E}$ scaling of the energy resolution of $\Kr$ events to higher energy events and is shown as the gray dashed lines. This uncertainty is based on previous NEXT-White energy resolution measurements with PMTs extrapolated from $\Kr$ energies to higher energies \cite{Next:krcal} compared with the measured energy resolution at higher energies \cite{NEXT:Eres}. This scaling uncertainty is likely an overestimate, as it is based on the PMTs, which have a poorer spatial uniformity than the SiPMs. 

With this extrapolation, for a high enough coverage, the contribution to the energy resolution from the photostatisics, as well as the SiPM noise, will be small, resulting in an excellent energy resolution, similar to that seen using PMTs in NEXT-White. With these conditions, the corrections will also be easier, as the issues with the response at low $z$ due to low coverage would be resolved.

\section{Discussion} \label{sec:dis}

By using the full SiPM waveforms of run 8088, a new method for measuring the energy in the NEXT-White detector is demonstrated. This new method allows for studying the energy resolution and response throughout the detector. An energy resolution of (10.9~$\pm$~0.6)~$\%$ (FWHM)  at \qty{41.5}{\kilo\electronvolt}, with a mean of (731.4~$\pm~0.3$)~PE and a standard deviation of (33.9~$\pm~0.3$)~PE is obtained using an optimized region of the detector. This was slightly larger than the expected resolution, given the known and measured contributions to the energy resolution, namely the Fano factor, photostatistics, and noise, calculated using Eq.~\ref{eq:eres_pe}. In particular, it is found that the noise of the SiPMs is not the leading contribution to the energy resolution, and can be easily measured with this method. The small difference in the estimate (9.68~$\pm$~0.01$\%$) and the measured (10.9~$\pm$~0.6~$\%$) energy resolution can be attributed to the difficulty of correcting a low coverage detector. As this analysis required dedicated runs to use the full SiPM waveforms, it is limited in statistics for improving the correction maps. 

The analysis used the full waveforms of the SiPMs, which is unrealistic for every run of detectors due to the heavy load on the DAQ and the large memory required. A possible solution for future data acquisition is to zero suppress a large portion of the waveform, and keep the full signal in the S2 region and a region far from S2 (with only noise). Future studies will be necessary to determine the required amount of non-zero suppressed waveforms when using SiPMs for an energy measurement. 

This energy resolution, scaled by $1/\sqrt{E}$, corresponds to ($1.42~\pm~0.44)\%$ (FWHM) at \qty{2458}{\kilo\electronvolt}, the $\Xe$ $Q_{\beta \beta}$ energy. High-energy events are also larger in both $x-y$ and $z$ (or time). This increased event volume will increase the number of SiPMs and time samples used to calculate the event energy, thus increasing the total noise of the event. The event volume is related to the event energy (or $N_{PE})$ as: $V \propto N_{PE}^{\gamma}$. The $\gamma$ exponent is bounded by $0 < \gamma < 1$. In the most conservative estimate of $\gamma = 1$, then $\sigma_n^2 \propto V \propto N_{PE}$. This would result in the noise term of the energy resolution from Eq.~\ref{eq:eres_pe} to scale as $\frac{\sigma_n^2}{N_{PE}^2} \propto \frac{1}{N_{PE}}$. Thus, the noise term, just like the photostatistics term, becomes smaller with larger energy events. High energy events likely have additional factors that are important in the energy resolution, for example, taking into account the more complex topology compared to point-like $\Kr$ events studied here. However, these effects should be bounded in the dashed lines of Figure~\ref{fig:est_eres} where the goal of sub-$1\%$ energy resolution is still achieved with appropriate coverage. 

By estimating the energy resolution as a function of coverage, it is shown that a NEXT-White-like detector with a higher coverage of SiPMs can achieve an energy resolution below 1\% (FWHM) at $Q_{\beta\beta}$. Therefore, for a future detector or prototype using only SiPMs for energy measurements, a higher coverage will be necessary.


\appendix

\section{Energy resolution with PMT signal} \label{sec:pmts}

\begin{figure}[htbp]
\centering 
\subfloat[]{\includegraphics[width=.47\textwidth]{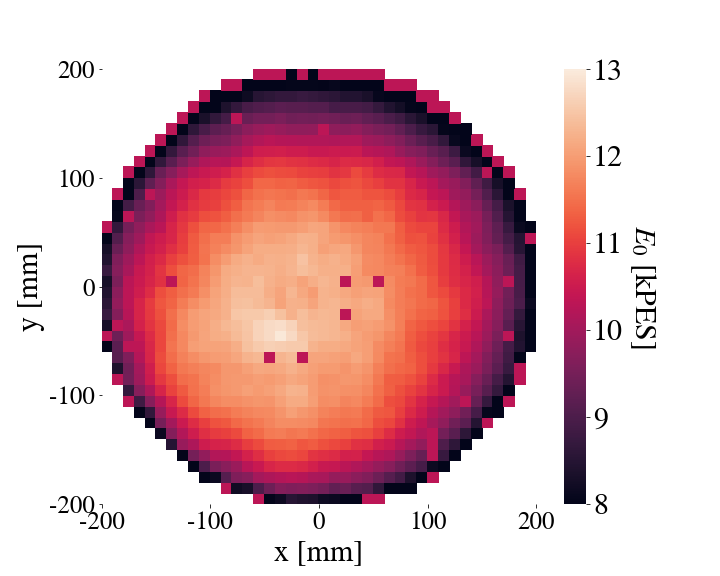}}
\subfloat[]{\includegraphics[width=.51\textwidth]{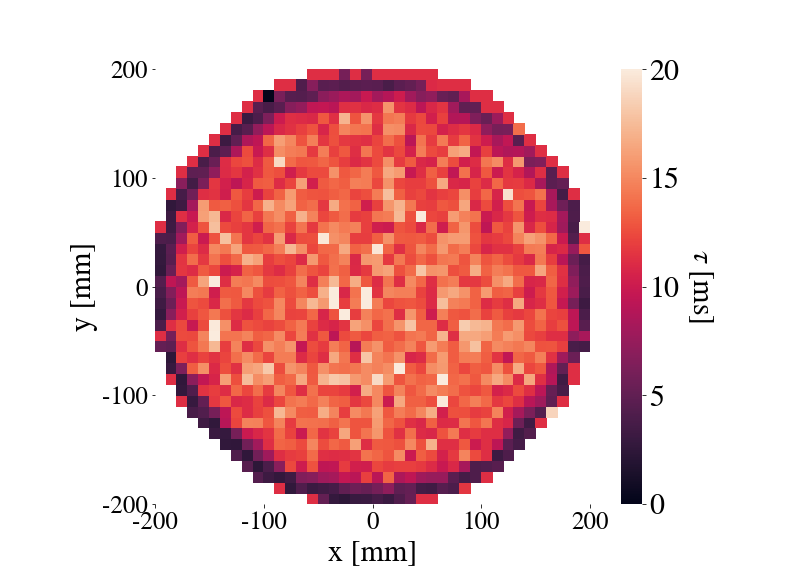}}
\caption{\label{fig:pmt_krmaps} Corrections maps using the calibration sample of run 8088 extracted from PMT signal, based on an exponential decay with (a) an initial energy at $z=0$ of $E_0$ and (b) lifetime $\tau$.}
\end{figure}

The $r$ and $z$ dependence of the energy resolution using the PMT signal is included here for comparison with the SiPMs throughout this analysis. The event energy seen by the PMTs is corrected using the calibration maps extracted from the PMT energy, shown in Figure~\ref{fig:pmt_krmaps}. The overall energy resolution using PMTs is consistent with the energy plane coverage and the previous low energy resolution measurements \cite{Next:krcal}.

\begin{figure}
    \centering
    \includegraphics[width=.6\linewidth]{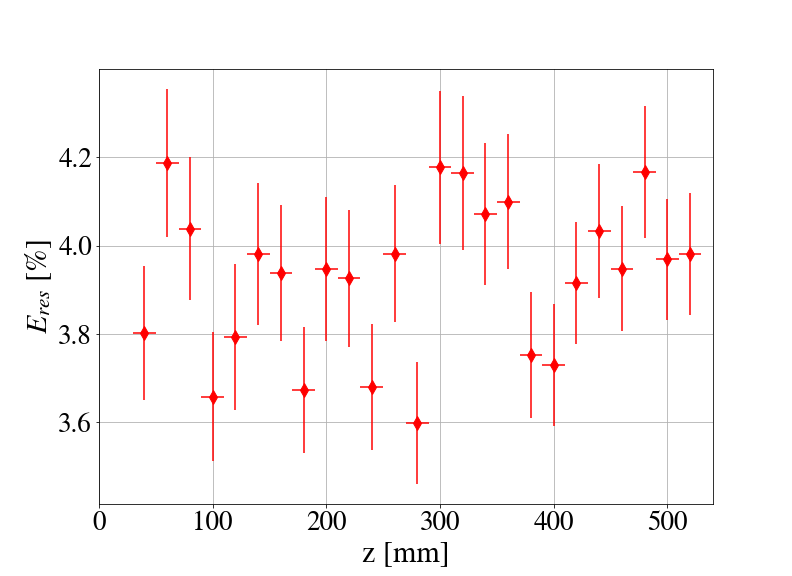}
    \caption{The energy resolution (FWHM) as a function of the $z$ position of events using PMTs, after corrections. Only events with $r<\qty{100}{\milli\meter}$ are included, to remove radial dependencies. The error bars correspond to statistical fluctuations.}
    \label{fig:z_eres_pmt}
\end{figure}

Figure~\ref{fig:z_eres_pmt} shows the energy resolution as a function of the $z$ position. There is a known trend of worsening energy resolution at higher $z$ (longer drift times), caused by the loss of energy from electron attachment to impurities in the gas. However, this effect is less prominent in Figure~\ref{fig:z_eres_pmt} than in previous NEXT analyses (\cite{Next:krcal}) due to the improvement in electron lifetime throughout the data taking of NEXT-White. The lifetime in \cite{Next:krcal} was \qty{2}{\milli\second}, while this run has a lifetime of $\sim$\qty{12}{\milli\second}. The worse energy resolution at very low $z$ when using the SiPMs signal, as seen in Figure~\ref{fig:z_vs_eres}, is not seen when using the PMT signal. This reinforces the conclusion that the worse energy resolution at low $z$ when using SiPMs is due to the geometry of the tracking plane. 

\begin{figure}
    \centering
    \includegraphics[width=.6\linewidth]{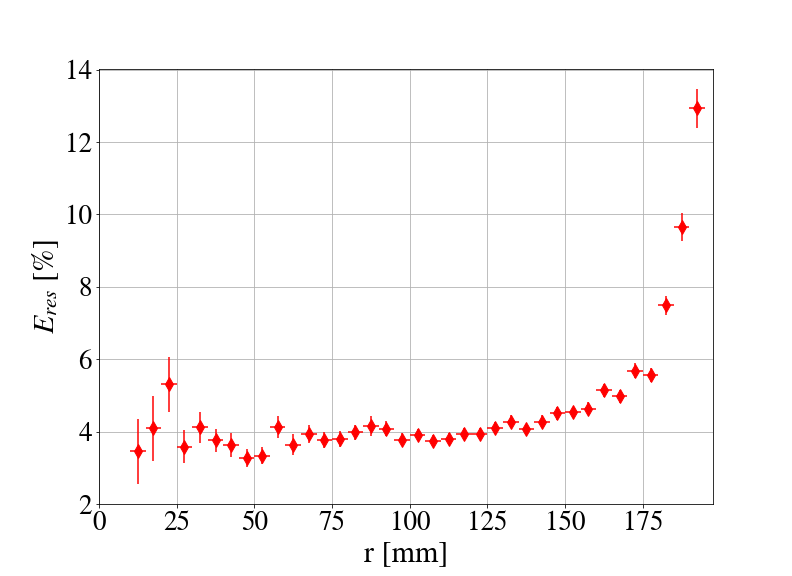}
    \caption{Energy resolution (FWHM) as a function of the radial distance using PMTs, after corrections. Only events with $\qty{50}{\milli\meter}<z<\qty{200}{\milli\meter}$ are included, to remove any longitudinal dependencies. The errors correspond to statistical fluctuations.}
    \label{fig:r_eres_pmt}
\end{figure}

Figure~\ref{fig:r_eres_pmt} shows energy resolution as a function of radial distance. The geometric effects of the detector have a large effect on the energy resolution when using PMTs, with larger $r$ resulting in a worse energy resolution. Since the energy plane is on the opposite side of the EL region, the light may reflect more to reach the PMTs compared to the SiPMs, causing the edge effects to be more significant.

\acknowledgments

The NEXT Collaboration acknowledges support from the following agencies and institutions: the European Research Council (ERC) under Grant Agreement No. 951281-BOLD; the European Union’s Framework Programme for Research and Innovation Horizon 2020 (2014–2020) under Grant Agreement No. 957202-HIDDEN; the MCIN/AEI of Spain and ERDF A way of making Europe under grants PID2021-125475NB and the Severo Ochoa Program grant CEX2018-000867-S; the Generalitat Valenciana of Spain under grants PROMETEO/2021/087 and CIDEGENT/2019/049; the Department of Education of the Basque Government of Spain under the predoctoral training program non-doctoral research personnel; the Spanish la Caixa Foundation (ID 100010434) under fellowship code LCF/BQ/PI22/11910019; the Portuguese FCT under project UID/FIS/04559/2020 to fund the activities of LIBPhys-UC; the Israel Science Foundation (ISF) under grant 1223/21; the Pazy Foundation (Israel) under grants 310/22, 315/19 and 465; the US Department of Energy under contracts number DE-AC02-06CH11357 (Argonne National Laboratory), DE-AC02-07CH11359 (Fermi National Accelerator Laboratory), DE-FG02-13ER42020 (Texas A\&M), DE-SC0019054 (Texas Arlington) and DE-SC0019223 (Texas Arlington); the US National Science Foundation under award number NSF CHE 2004111; the Robert A Welch Foundation under award number Y-2031-20200401. Finally, we are grateful to the Laboratorio Subterr\'aneo de Canfranc for hosting and supporting the NEXT experiment.


\bibliographystyle{JHEP}
\bibliography{biblio.bib}






\end{document}